\documentclass[floatfix,aps, preprint,prb,superscriptaddress, amsfonts]{revtex4}

\usepackage{graphicx}

\usepackage{dcolumn}

\usepackage{natbib}

\usepackage{bm}

\begin{document}

\title{Ellipsometric study of the Electronic Structure of Ga$_{1-x}$Mn$_x$As and LT-GaAs}

\author{K.S.~Burch}

\email{burch@physics.ucsd.edu}

\affiliation{Department of Physics, University of California, San Diego, CA 92093-0319}

\author{J.~Stephens}

\affiliation{Center for Spintronics and Quantum Computation, University of California, Santa Barbara, CA 93106}

\author{R.~K.~Kawakami}

\altaffiliation[Permanent address: ]{Department of Physics, University of California, Riverside, CA 92521}

\affiliation{Center for Spintronics and Quantum Computation, University of California, Santa Barbara, CA 93106}

\author{D.D.~Awschalom}

\affiliation{Center for Spintronics and Quantum Computation, University of California, Santa Barbara, CA 93106}

\author{D.N.~Basov}
\affiliation{Department of Physics, University of California, San Diego, CA 92093-0319}

\begin{abstract}
We have measured the optical constants of Ga$_{1-x}$Mn$_{x}$As from 0.62 eV to 6 eV, using spectroscopic ellipsometry. The second derivatives of the dielectric function are examined through a critical point analysis. The $E_{1}$ critical point shifts to higher energies with increased doping of Mn, while all other critical points appear unaffected. The evolution of the critical points results from the interplay between band gap renormalization from ionized impurities and sp-d hybridization of the Mn induced impurity band and GaAs valence and conductions bands.

\end{abstract}

\maketitle

\section{Introduction}
\label{sec:intro}
Semiconductors doped with magnetic impurities, generally referred
to as Diluted Magnetic Semiconductors (DMS), have produced great scientific
and technological interest in recent years.\cite{ohno} Such systems
offer a promising opportunity to explore devices that simultaneously exploit the
spin and charge degrees of freedom.\cite{Wolf01} They also bring the challenge
of understanding the physics involved in the coupling of local moments in d orbitals with sp extended states. One of the most widely studied DMS is Ga$_{1-x}$Mn$_{x}$As,
in part because GaAs is a well characterized semiconductor used  in a variety of digital signal processing circuits, telecommunication systems, and optoelectronics. While there is general agreement that ferromagnetism in Ga$_{1-x}$Mn$_{x}$As is driven by a carrier mediated mechanism between the local moments (Mn 3d$^{5}$ electrons) and the carriers
(holes),\cite{2fluidsl} the evolution of the electronic structure
with Mn doping as well as it's role in the ferromagnetism is still
under debate. 

\begin{figure}
\includegraphics*[width=6.0in]{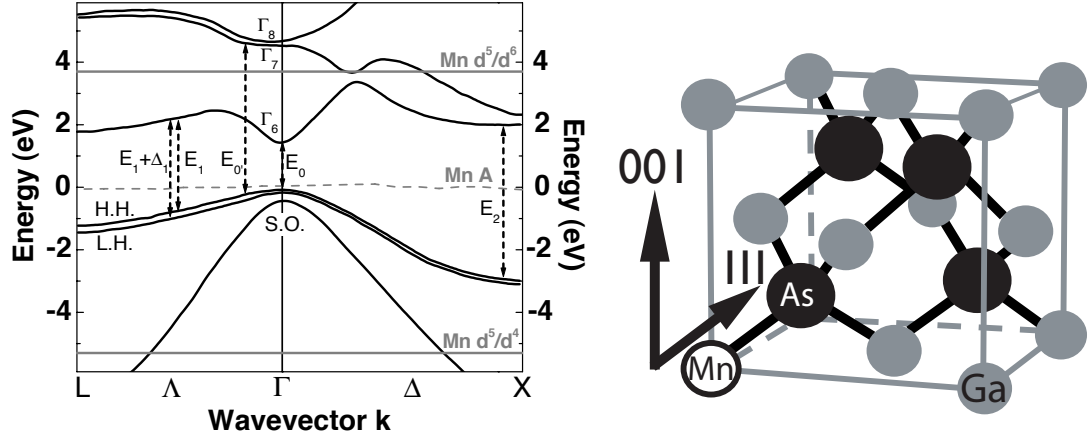}
\caption{\label{fig:bndstrc} Left: GaAs band structure and relevant critical point transitions reproduced from Ref. \onlinecite{Cardona1}. The upper conduction bands are labeled as $\Gamma_{7,8}$ based on symmetry, while the lowest conduction band is labeled $\Gamma_{6}$. The valence bands have been labeled as H.H. for heavy-hole, L.H. for light-hole, and S.O. for split-off. Taken from Ref. \onlinecite{Fujimori}, Mn d filled (d$^5$/d$^4$) and empty (d$^5$/d$^6$) levels are shown in grey, and the acceptor Mn A is dashed-gray. The dispersion of the Mn acceptor level is also taken from Ref. \onlinecite{Fujimori}. The L point corresponds to the 111 direction and the X point to the 001 direction. Right: The Ga$_{1-x}$Mn$_{x}$As unit cell with the important symmetry directions labeled.}
\end{figure}

The controversy around the electronic structure of Ga$_{1-x}$Mn$_{x}$As
generally centers around the position of the Fermi level. One picture places the holes in the mn induced impurity band\cite{Dagotto, Berciu, DasSarma}, while others place the Fermi level in an unperturbed GaAs valence band.\cite{Macdonald, 2fluidsl, Sinova} These differing viewpoints are in part driven
by the early work of Ohno et. al., who showed the onset of ferromagnetic
behavior in Ga$_{1-x}$Mn$_{x}$As with increased doping, was at or near the Metal
to Insulator transition.\cite{ohno} Additionally, optical absorption measurements
established the formation of a Mn induced shallow acceptor level 110
meV above the valence band in paramagnetic GaAs doped with Mn in the
very dilute limit.\cite{gamnasoptics} Recent STS and ARPES experiments suggest the Mn form an "impurity band" of d-like states that strongly hybridize with the GaAs valence band.\cite{STS, Fujimori, mathieu} The ARPES measurements place the occupied d$^{5}/d^{4}$ levels $\approx 5.3~eV$ below the valence band maximum (VBM), with the unoccupied d$^{5}/d^{6}$ level 3.7 eV above the VBM (see Fig. \ref{fig:bndstrc}).\cite{Fujimori} Nonetheless these measurements are limited in resolution and scope, and therefore the quantitative evolution of the band structure with x has yet to be established experimentally.  Infrared spectroscopy measurements established the role of this impurity band in the carrier dynamics of Ga$_{1-x}$Mn$_{x}$As, however they were limited to energies below the band gap and could only discuss effects at the zone center.\cite{jasonpapers,burch} 

Experimental studies of the Ga$_{1-x}$Mn$_{x}$As electronic structure that combine high resolution, broad doping range, and do not focus on the zone center are needed to address several key issues. Interestingly, although Ga$_{1-x}$Mn$_{x}$As is generally referred to as an "alloy", implying the momentum ($\overrightarrow{k}$) is conserved and is still a good quantum number, this has yet to be confirmed experimentally. Additionally the effects on the GaAs band structure of sp-d hybridization between the Mn d and As/Ga sp states are still unknown. Finally the spin-orbit splitting in Ga$_{1-x}$Mn$_{x}$As has yet to be measured, despite being critical to the usefulness of Ga$_{1-x}$Mn$_{x}$As as a spintronic device. To investigate these and other effects of Mn doping we have a performed a line shape analysis of the complex dielectric function determined by spectroscopic ellipsometry.

For the past four decades, spectroscopic ellipsometry has provided key insights into the electronic structure of many materials.\cite{Cardona4} Unlike common spectroscopic techniques, ellipsometry measures the amplitude ($\Psi$) and phase ($\Delta$) of the reflected wave. Therefore the complex dielectric response ($\hat{\epsilon}$) of a material can be obtained analytically in bulk materials.  Using standard techniques the optical constants of a layered structure can be determined with high resolution over a broad energy range. Strong features in the spectra  result from interband transitions at different points in the Brillouin zone (see Fig. \ref{fig:bndstrc}). A critical point analysis of $\hat{\epsilon}$ provides direct determination of the subtle features that can be compared with band structure calculations. This motivated us to perform an ellipsometric study of Ga$_{1-x}$Mn$_{x}$As such that a detailed picture of the evolution of the band structure at a number of points in k-space can emerge. Similar efforts on II-VI DMS have aided in the determination of the strength of sp-d hybridization (V$_{(s,p)d}$) in these materials.\cite{Znelip} An accurate understanding of the role of sp-d hybridization in DMS is critical, as a strong V$_{(s,p)d}$ can lead to the formation of a Zhang-Rice polaron, binding the Mn induced hole. The strength of V$_{(s,p)d}$ will also determine the strength of the hopping amplitude "t" of the holes,\cite{ZR,Dietl} central to a number of different theories of ferromagnetism in Ga$_{1-x}$Mn$_{x}$As.\cite{Dagotto,Berciu,DasSarma} Additionally, the kinetic exchange, which plays a large role in the magneto-optical properties of Ga$_{1-x}$Mn$_{x}$As,\cite{Sinova} can be related to the sp-d hybridization via second order perturbation theory (N$_{0}\beta \propto V^{2}_{pd}$).\cite{Hass} As discussed in Sec. \ref{sec:perturb}, sp-d hybridization will also result in sp bands avoiding the Mn d levels,\cite{hybrid,Znelip,cdelip,Hass} and is therefore central to understanding the evolution of the band structure in Ga$_{1-x}$Mn$_{x}$As. 

Our spectroscopic investigation has uncovered the evolution of band structure of mn doped GaAs. Specifically,  from the critical point analysis we clearly uncover the important role of hybridization between Mn induced impurity band and the GaAs valence band. Namely the anisotropic strength of this hybridization results in a blue shift of the $E_{1}$ transition while all other critical points remain unchanged. We would also like to note that at room temperature the $E_{0^{'}}$ and $E_{2}$ critical points (see Fig. \ref{fig:bndstrc}) are the mixture of two or more critical points where $\Delta E_{g} \leq \Gamma$.\cite{Cardona3} The analysis as well as it's results are discussed in Sec. \ref{sec:cpa}. The measured elipsometric data can be found in Sec. \ref{sec:pd}. The samples and experimental methods are described in Sec. \ref{sec:sample}. The fitting procedure and the dielectric function is detailed in Sec. \ref{sec:model}. Finally we discuss the implications for each critical point in Sec. \ref{sec:discuss}

\section{Samples and Experimental Method}
\label{sec:sample}
	The samples in this study were grown at UCSB on semi-insulating GaAs (100) by low temperature molecular beam epitaxy (LT-MBE). The Ga$_{1-x}$Mn$_{x}$As and LT-GaAs samples were deposited at a temperature of $260^{\circ}\mathrm{C}$. The sample labeled GaAs is a bare substrate. The Ga$_{1-x}$Mn$_{x}$As layers had a nominal thickness of 500nm and were grown atop a 60 nm LT-GaAs buffer layer. The LT-GaAs sample had a nominal thickness of 1500 nm (see Fig. \ref{fig:model} for details).  The oxide and buffer layers were taken into account using a multiphase analysis described below. 
	
	Spectroscopic ellipsometry $(0.62-6~eV)$ and near-normal transmission  (T) over the energy range $0.005\rightarrow 1.42~eV$ measurements were performed at UCSD at room temperature. Details of the transmission measurements can be found in Ref. \onlinecite{jasonpapers}. For the ellipsometry experiments the back surface of the substrate was roughened so as to prevent interference in the substrate. A variable angle spectroscopic ellipsometer (VASE) instrument from J. A. Woollam and Associates with a rotating analyzer and an auto-compensator measured the complex ellipsometric ratio ($\rho$) at 65$^{\circ}$ and 75$^{\circ}$ angle of incidence. $\rho$ is the ratio of the reflectance coefficients $r_p$ and $r_s$ (parallel and perpendicular to the plane of incidence). This is generally expressed in terms of two angles $\Psi$ and $\Delta$: 
	\begin{equation}
	 \rho=\frac{r_{p}}{r_{s}}=e^{i \Delta} tan\Psi 
	\label{eq:ratio} \end{equation}
where $\Psi$ is a measure of the relative amplitude and $\Delta$ the relative phase shift. From $\Psi$ and $\Delta$ the complex dielectric function ($\hat{\epsilon}=\epsilon_{1}+i\epsilon_{2}$) can be readily derived using the two-phase model ($ambient+sample$).\cite{Azzam} In real materials surface roughness, oxide overlayers and the multilayered nature of the samples provides a situation for which no analytic solution currently exists. However, genuine optical constants can be obtained through the use of a multiphase model.\cite{Azzam} A significant parameter in evaluating these models is the penetration depth of the incident light ($\delta$):
\begin{equation}
	\label{eq:depth} \delta=\frac{\lambda}{4\pi k}
	 \end{equation}
where $\lambda$ is the wavelength of the incident light and $k$ is the complex part of the index of refraction ($\sqrt{\hat{\epsilon}}=\hat{n}=n+ik$). If a layer has a thickness greater than $2\delta$ then the light from layers below it do not contribute to the measured $\Psi$ and $\Delta$, as it is attenuated 50 times.\cite{Cardona1} Therefore in regions where $\epsilon_{2}$ is large and/or at higher energies, the primary contribution is from the top few atomic layers. Specifically in the region of the $E_{1}$ critical point $\delta(E_{1})\approx20~nm$, whereas near $E_{2}$ $\delta(E_{2})\approx5~nm$.
	
\section{Results and Analysis}
\label{sec:RA}
	\subsection{$\Psi$ and $\Delta$}
	\label{sec:pd}
\begin{figure}
\includegraphics*[width=5.5in]{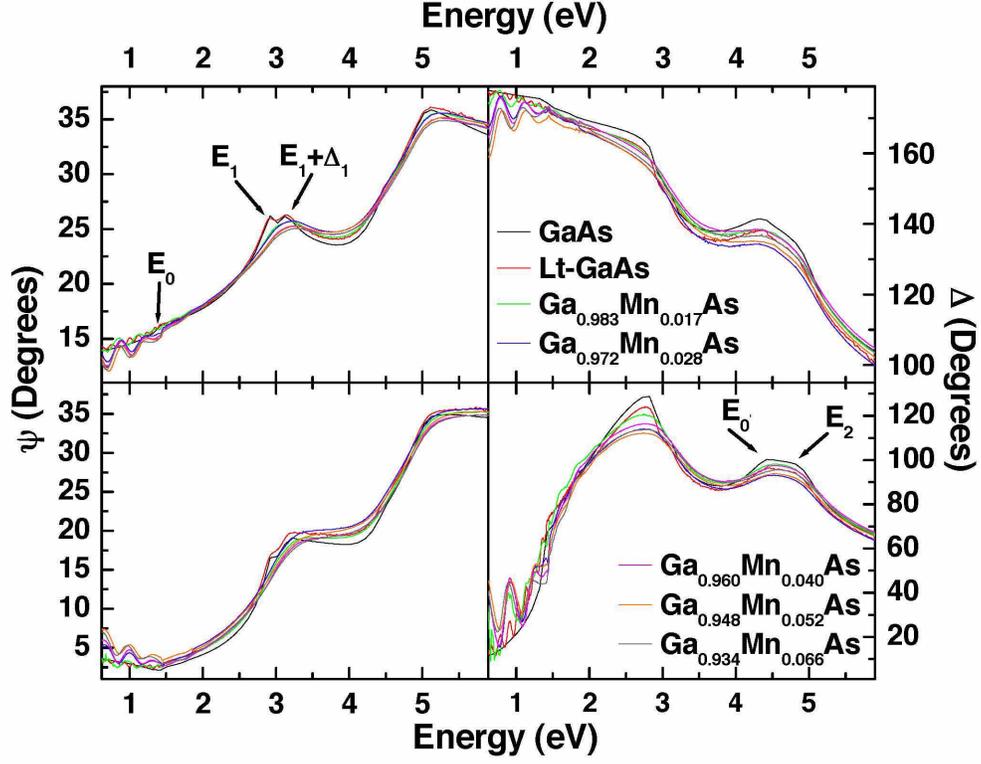}
\caption{\label{fig:psidel}Ellipsometric angles $\Psi~\&~\Delta$ measured at a 65$^{\circ}$ (top panels) and 75$^{\circ}$ (bottom panels) angle of incidence. The interference fringes at low energies are due to interference from the thin film. The two peaks around 3 eV are due to the $E_{1}~\&~E_{1}+\Delta_{1}$ critical points, which clearly broaden and blue shift with Mn doping. However,  Mn doping has little effect on  the two extremum around 4.5 \& 5 eV are due to the $E_{0^{'}}~\&~E_{2}$ critical points.}
\end{figure}

	In Fig. \ref{fig:psidel} we plot the measured ellipsometric parameters at  65$^{\circ}$ (top panels) and 75$^{\circ}$ (bottom panels) angle of incidence respectively. We first take note of the significant difference in the shape and magnitude of $\Delta$ at these two angles. The uniqueness of the information garnered at the measured angles is the result of taking data just below and above the Brewster's angle for GaAs. Turning our attention to the low energy portion of the spectra ($E\leq1.75eV$), interference fringes appear in all samples except the bare substrate. In this range we approach the fundamental band gap of GaAs,  which can be seen as sharp points around 1.42eV in both  $\Psi$ and $\Delta$. Furthermore, in this region k becomes sufficiently small and $\lambda$ adequately long so that $2\delta$ is greater than the thickness of the deposited film. Therefore the strength and position of these fringes provides important additional information about the thickness of the film as well as it's optical constants. \cite{backsub}  
	
	We now examine the region between 2.5 and 3 eV. Focusing first on $\Psi$, we see that at both angles the GaAs data displays two sharps points. These are the $E_1$ and $E_1+\Delta_1$ critical points to be discussed in subsection \ref{sec:E1} (see Fig. \ref{fig:bndstrc}). These critical points are broadened in the LT-GaAs sample and in all of the Ga$_{1-x}$Mn$_x$As samples they appear to have merged. This trend can also be seen in the $\Delta$ data taken at $75^{\circ}$ (see Fig. \ref{fig:psidel}). In Fig. \ref{fig:psidel} we note a reduction in $\Psi$ between 2.75 eV and 3 eV and concurrent growth below 2.5eV. 
	
	Finally we turn our attention to the region between 4 eV and 5 eV. While data in this region is affected by the native oxide layer, discussed further in Sec. \ref{sec:E2}, there is an important trend worth noting. This is best seen in $\Psi$ at $75^{\circ}$, where ellipsometry are evident in the GaAs data. While the sharpness of the peaks appears reduced in the LT-GaAs and Ga$_{1-x}$Mn$_{x}$As samples, this does not seem to be the result of significant broadening. Most notably the position of these two peaks remains unchanged with Mn doping. 
	\subsection{Modeling the Optical Constants}
	\label{sec:model}
\begin{figure}
\includegraphics*[width=2.0in]{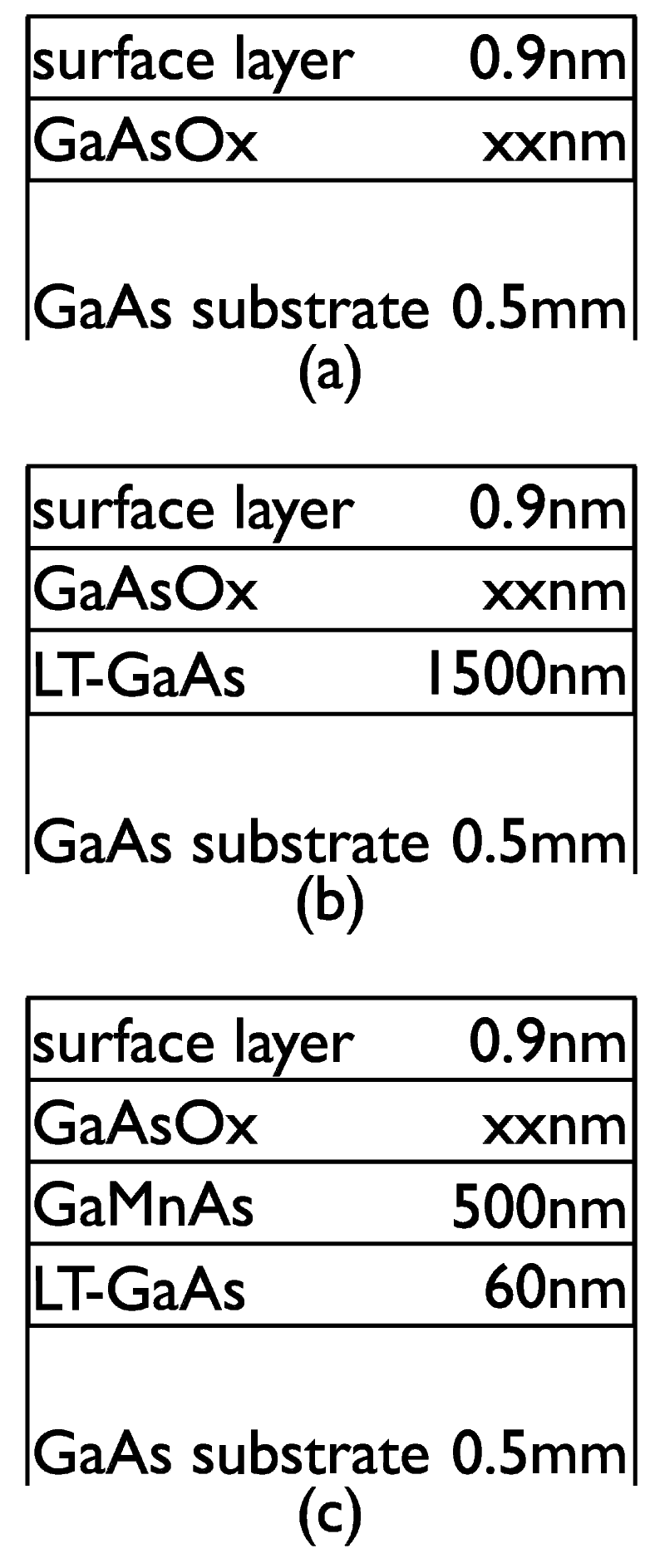}
\caption{\label{fig:model} A schematic diagram of (a) the GaAs substrate model, (b) the model for the LT-GaAs sample, and (c) the Ga$_{1-x}$Mn$_{x}$As samples.}
\end{figure}

	As noted in Sec. \ref{sec:sample}, the optical constants cannot be obtained analytically for any of the samples in this study due to surface roughness and the presence of an oxide layer.\cite{Aspnes1} This problem is compounded by the multilayered nature of the samples. Therefore to obtain the optical constants of the films we have devised a method to properly model these samples. To simplify this problem we first measured $\Psi$ and $\Delta$ for a piece of GaAs substrate, which had approximately the same exposure to air and roughening conditions as the other samples in this study. The substrate was successfully modeled with three layers (see Fig. \ref{fig:model} a). The first contained the known optical constants of GaAs with a fixed thickness of 0.5mm. The next two layers were GaAsOx (Native oxide), and a surface layer modeling roughness as an effective medium of 50\% void and 50\% GaAsOx (see Fig. \ref{fig:model}).\cite{Aspnes1,jellison} We then performed a least squares fit to $\Psi$ and $\Delta$ to determine the oxide and surface layer thicknesses. 
	
	Next we modeled the Lt-GaAs data similar to GaAs with an additional 1500 nm thick layer between the substrate and the oxide layer (see Fig. \ref{fig:model} b). Initially the thickness of the oxide and surface layers were the same as those determined for the substrate.  The optical constants of the LT-GaAs layer were defined using a sum of Lorentzian and Tauc-Lorentzian oscillators:
	\begin{equation}
	\label{eq:epsilon}
\hat{{\epsilon}}=\epsilon_{0}+\sum_{i}\hat{{\epsilon}}_{i}^{Lorentz}+\sum_{j}\hat{{\epsilon}}_{j}^{Tauc-Lorentz} 
	\end{equation}
	\begin{equation}
	\label{eq:lorentz} \hat{{\epsilon}}_{i}^{Lorentz}=\frac{A_{i}\Gamma_{i}E_{i}}{E_{i}^{2}-E^{2}-i\Gamma_{i}E} 
	\end{equation}
	\begin{equation}
	\label{eq:tauc}  \hat{{\epsilon}}_{j}^{Tauc-Lorentz}=\frac{2}{\pi}P\int_{E_{bi}}^{\infty}\frac{\zeta}{\zeta^{2}-E^{2}}\frac{A_{i}(\zeta-E_{bi})^{2}}{(\zeta^{2}-E_{ci}^{2})+i\Gamma_{i}^{2}}d\zeta+i\left[\frac{A_{i}(E-E_{bi})^{2}}{(E^{2}-E_{ci}^{2})+i\Gamma_{i}^{2}}\frac{\Theta(E-E_{bi})}{E}\right] 
	\end{equation}
	where $\Theta(E-E_{bi})$ is the unit step function, P implies the Cauchy principle value, and $\epsilon_{0}$ is a constant used to model the polarizability of the material. Three Lorentzian oscillators where employed to model the effects of one and two phonon absorption in the infrared portion of the spectrum.\cite{Wooten} The Tauc-Lorentzian oscillators, see eq. \ref{eq:tauc}, were utilized to model the effect of interband transitions.\cite{Jellison} We note that we choose to model the optical constants using oscillators instead of performing a least squares fit for $\hat{\epsilon}$ directly so as to ensure the results are Kramers-Kronig consistent. This approach also enabled us to improve upon standard techniques by including transmission data and the effect of oscillators centered below the ellipsometer's range. Lastly we note that for $0.62~eV\leq E\leq1.42~eV$ this procedure produced optical constants consistent (within $1\%$) with previous results derived from a combination of normal incidence transmission and reflection.\cite{jasonpapers} 
	
	To obtain the initial conditions for the LT-GaAs generic layer, we first fit the optical constants of GaAs using eq. \ref{eq:epsilon}. We then applied this model to the LT-GaAs data and performed a fit for the thicknesses of the LT-GaAs, oxide and surface layers. Next we fit for the parameters of each oscillator separately. This was done to avoid the effect of correlations due to the large overlap of the oscillators. Once all the oscillators had been fit, we refit the thickness of each layer. This iterative method was performed until the fit could no longer be improved. We repeated the fitting procedure with a number of different initial conditions so as to ensure the final answer was not dependent on our original values. 
	
	The Ga$_{0.983}$Mn$_{0.017}$As data was fit after the LT-GaAs sample, using a similar approach, however the model now contained a 500nm Ga$_{0.983}$Mn$_{0.017}$As layer atop a 60nm LT-GaAs layer (see Fig. \ref{fig:model}). Since the penetration depth for most of the fitted range was less than 500nm, the thickness of the LT-GaAs layer was never allowed to vary due to it's weak contribution to the data. The remaining Ga$_{1-x}$Mn$_{x}$As samples were fit in a similar fashion, however they contained two additional oscillators. The first modeled the effect of free carriers using the Drude form (a Lorentzian with $E_{i}=0$), and the second was an additional Tauc-Lorentzian oscillator to model the effect of interband transitions from the GaAs valence band to the Mn induced impurity band.	

	The $\hat{\epsilon}$ resulting from the modeling can be seen in Fig. \ref{fig:e1e2}. The critical points of GaAs have been labeled in the graph of $\epsilon_2$. Consistent with our earlier work on these samples, we find that the fundamental band gap ($E_{0}$) is "smeared" out in LT-GaAs and Ga$_{1-x}$Mn$_{x}$As samples.\cite{jasonpapers} We note that this effect can be seen in both $\epsilon_1$ and $\epsilon_2$. The origin of this smearing will be discussed in Sec. \ref{sec:E0}, however Fig. \ref{fig:e1e2} demonstrates that this broadening grows with Mn doping until $x=0.028$. Additionally this effect seems to extend to $\sim2.75~eV$.  This smearing appears to be aided by a transfer of spectral weight from the region between $2.75~eV$ and $3.25~eV$ to the region below 2.75~eV. 
	
	We now discuss the region of the $E_{1}$ and $E_{1}+\Delta_{1}$ transitions, namely $2.5~eV\rightarrow 3.5~eV$. First focusing on $\epsilon_1$ we note that as we go through the series the peak at 2.85~eV broadens and decreases in strength. Turning our attention to $\epsilon_2$ we see that mainly the $E_1$ peak is broadened and decreases in strength in LT-GaAs and appears to disappear in the Mn doped samples. While the broadening and reduction in amplitude is consistent with previous studies of doped GaAs, these works revealed a red shifting of both the $E_{1}$ and $E_{1}+\Delta_{1}$ transitions, whereas we observe a blue shifting.\cite{Cardona2} Additionally, in Ga$_{1-x}$Mn$_{x}$As the two peaks appear to merge. As we discuss in sec. \ref{sec:E1}, this merging is the combined result of increased broadening and sp-d hybridization. 
	
	Finally we turn our attention to the region of the $E_{0^{'}}$ and $E_{2}$ critical points ($4.25eV\rightarrow5.25eV$) in Fig. \ref{fig:e1e2}. Despite the presence of the oxide layer and the small penetration depth ($\delta\approx 5~nm$), the critical points can still be clearly recognized in all $\epsilon_{1}$ spectra and in most of the $\epsilon_{2}$ spectra. Focusing on $\epsilon_{1}$, we see that the position and broadening of the critical points appears almost constant throughout the series. Not surprisingly, the amplitude of this peak appears to be random, as previous elipsometric studies established the effect of the oxide layer reduces the strength of the measured $E_{2}$ peak.\cite{Aspnes2} Therefore we do not expect the presence of the Oxide layer to significantly effect our analysis.
		\subsection{Critical Point Analysis}
		\label{sec:cpa} 
\begin{figure}
\includegraphics*[width=5.5in]{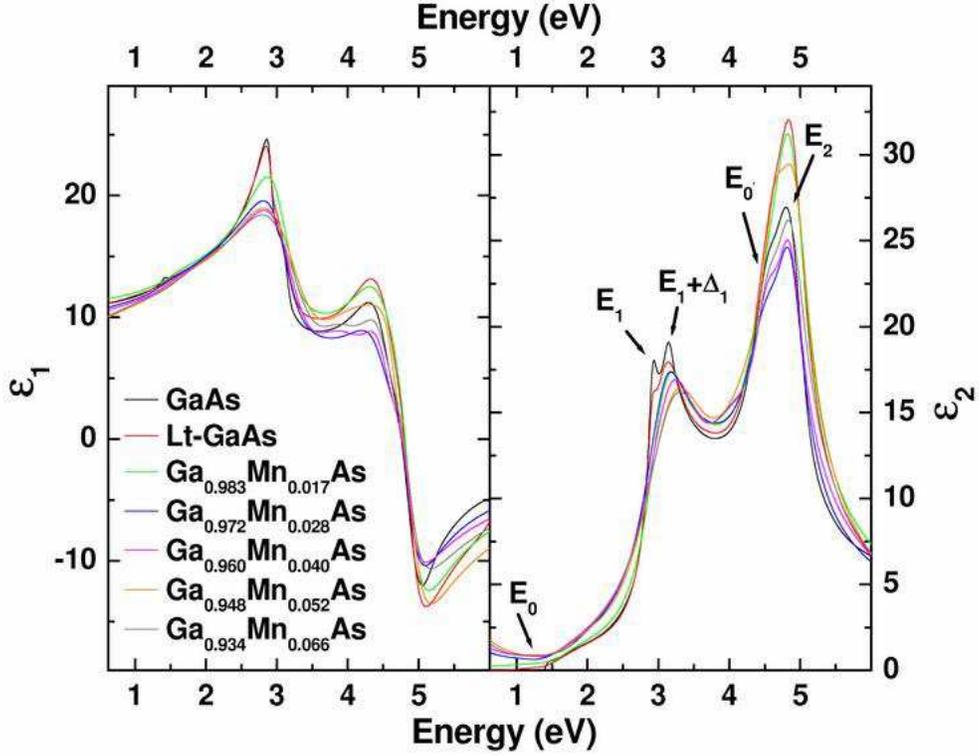}
\caption{\label{fig:e1e2} Left panel: The real part of the dielectric function for all samples in this study. Right panel: The imaginary (absorptive) part of the dielectric function with the critical points labeled. In both panels we clearly see the broadening of $E_{0}~\&~E_{1}$ with Mn doping, while the right panel clearly demonstrates the blue shifting of $E_{1}$. We also note the apparent lack of change in $E_{0^{'}}~\&~E_{2}$.  }
\end{figure}

\begin{figure}
\includegraphics*[width=7.0in]{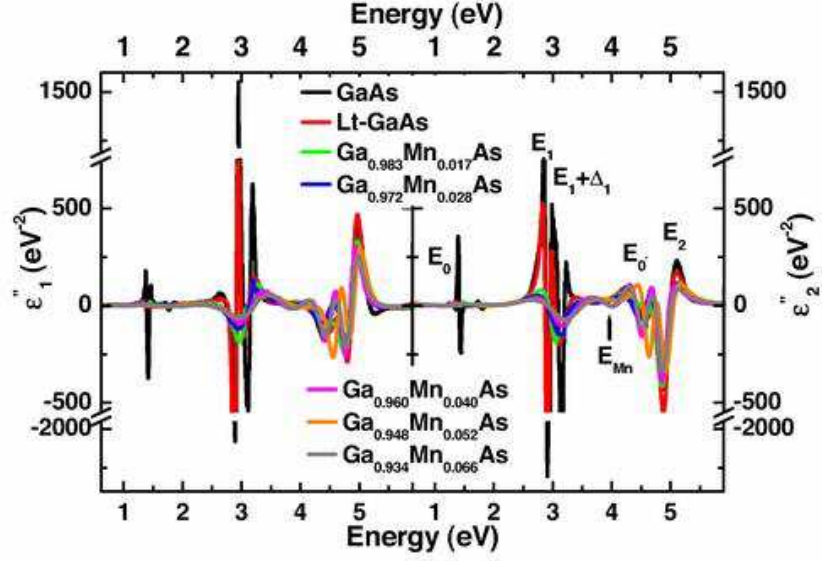}
\caption{\label{fig:fullderivspec}The derivative spectra of all samples in this study, which allow a clear identification of all critical points. We note the complete loss of a feature at $E_{0}$ in all sample grown bylow temperature MBE. The $E_{1}$ critical point is significantly broadened and blue shifted with Mn doping, while $E_{0^{'}}~\&~E_{2}$ show little change.}
\end{figure}

		The numerical second derivatives of the $\hat{\epsilon}$ data presented in Fig. \ref{fig:e1e2} can be found in Fig. \ref{fig:fullderivspec}. A cursory examination of this graph quickly reveals it's utility in analyzing the structures seen in the $\hat{\epsilon}$ spectra. Before discussing the results separately for each of the relevant critical points, we briefly mention some general trends in the data. The $E_{0^{'}}$ \& $E_2$ Critical Points, with the exception of the Ga$_{0.948}$Mn$_{0.052}$As sample, appear almost completely unaffected by growth at low temperature and/or Mn doping. We believe the anomalous behavior of the Ga$_{0.948}$Mn$_{0.052}$As sample results from having had the longest exposure to air (see Tbl. \ref{TBL}), however it's origin is not entirely clear. Interestingly for samples with $x\geq0.04$, an extremely weak extra feature (labeled E$_{Mn}$) appears at energies just below $E_{0^{'}}$. The origin of this peak will be discussed in Sec. \ref{sec:E0'}. 
		
		 Next we turn our attention to the $E_{0}$ and $E_{0}+\Delta_{0}$ transitions, which undergo a substantial change attributable to the low temperature growth. Namely, these transitions are no longer observable in the $\frac{d^{2}\hat{\epsilon}}{dE^{2}}$ spectra and therefore we have not attempted to fit these transitions in any sample, with the exception of the GaAs substrate. However, given the band edge broadening seen in Fig. \ref{fig:e1e2}, this result is not surprising. 
		
\begin{figure}
\includegraphics*[width=5.0in]{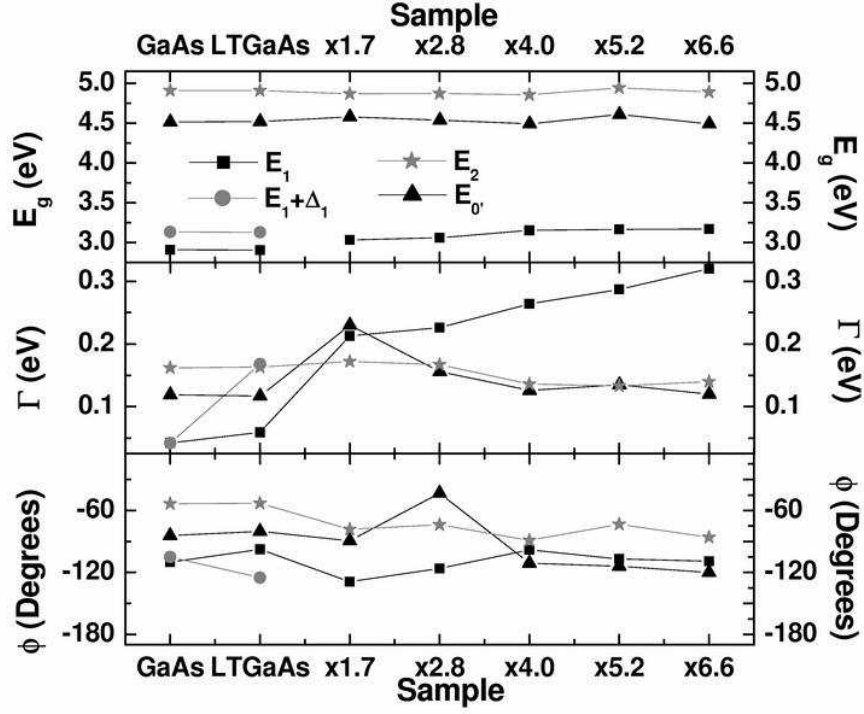}
\caption{\label{fig:fitparams} (Top panel) The resonant energy of each critical point for all samples. We note the increase in $E_{1}$ with increasing x, while all other points remain unchanged. (Middle Panel)The broadening of the critical points for each sample. The sudden change in the x=1.7$\%$ sample is due to the merging of $E_{1}~\&~E_{1}+\Delta_{1}$. (Bottom panel) The phenomenological phase parameter which accounts for the mixing of different critical points due to coulomb effects. Lines are guides to the eyes.} 
\end{figure}

		Let us now examine the $E_{1}$ and $E_{1}+\Delta_{1}$ critical points, which contain rather surprising results. We begin by comparing LT-GaAs and GaAs, noting a significant reduction in the amplitude of the critical points in the former with respect to the later. However in LT-GaAs the broadening of the $E_{1}$ critical point appears unchanged by low-temperature growth while the  $E_{1}+\Delta_{1}$ appears to be significantly broadened. As we expect from Fig. \ref{fig:e1e2}, the effect of Mn doping is quite dramatic. In all Mn doped samples, the broadening of the  $E_{1}$ and $E_{1}+\Delta_{1}$ critical points is such that they appear to merge. Additionally this merged structure is continuously blue-shifted as x is increased. When the $E_{1}$ structure just overlaps the $E_{0^{'}}$ critical point, it results in $E_{0^{'}}$ appearing more asymmetric. We therefore conclude that the significant broadening and blue shifting of these critical points is responsible for the apparent anomalies at x=.017,0.028  in Figs. \ref{fig:fitparams} . In the samples with higher dopings, the amplitude of the $E_{1}$ critical point continues to be reduced and the overlap between $E_{1}$ and $E_{0^{'}}$ increases, reducing the asymmetric effect of $E_{1}$ on $E_{0^{'}}$. 
In GaAs at room temperature, the derivative spectra in the vicinity of a critical point are well characterized by two-dimensional line shapes\cite{Cardona3,Aspnes2}:
\begin{equation}
\label{eq:2dcp} \frac{d^{2}\epsilon}{dE^{2}}=Ae^{i\Theta}(E-E_{g}+i\Gamma)^{-2} 
\end{equation}
where A is the amplitude of the critical point related to the reduced effective mass of the two bands involved in the transition, $E_{g}$ is the energy of the critical point and  $\Gamma$ is a broadening parameter determined by the quasiparticle lifetime and the  relaxation of the requirement of momentum conservation. The phenomenological parameter $\Theta$ is added to account for coulomb and excitonic effects that result in the admixture of two critical points.\cite{Aspnes3} The mixture of a minimum and a saddle point corresponds to $0\leq \Theta \leq \frac{\pi}{2}$, whereas the combination of a saddle point and a maximum corresponds to $\frac{\pi}{2} \leq \Theta \leq \pi$. 

\begin{figure}
\includegraphics*[width=5.5in]{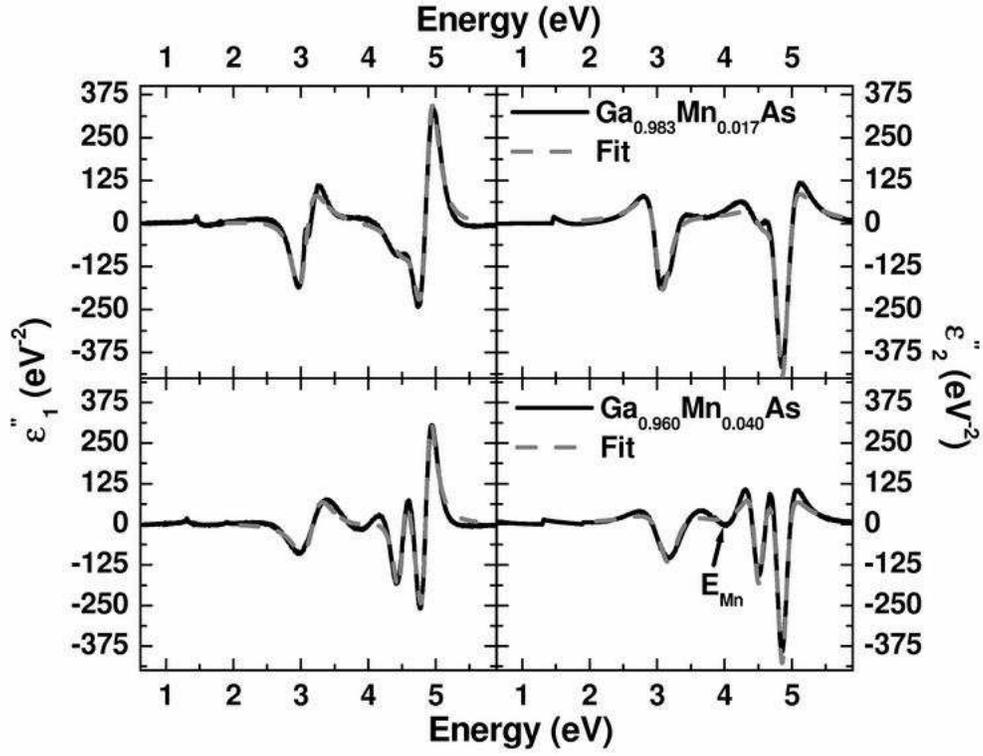}
\caption{\label{fig:fits} Two representative fits of $\frac{d^{2}\epsilon}{dE^{2}}$. In the bottom panel the extra feature at $E\approx4.0~eV$ can be seen, however it is too weak to provide a reliable fit.}
\end{figure}

Two representative plots of the $\frac{d^{2}\hat{\epsilon}}{dE^{2}}$ spectra generated by least-squares fitting are compared to the experimental results in Fig. \ref{fig:fits}. We started the 2D line shape analysis with GaAs and LT-GaAs. In GaAs and LT-GaAs the $E_{1}$ and $E_{1}+\Delta_{1}$ critical points were fit simultaneously assuming a constant spin orbit splitting ($\Delta_{1}=.224eV$). The  $E_{0^{'}}$ \& $E_{2}$ critical points were also fit together, however constant separation between the two was not assumed.\cite{Cardona1} Since we were unable to distinguish the  $E_{1}+\Delta_{1}$ critical point from $E_{1}$ in  the Ga$_{0.983}$Mn$_{0.017}$As sample, we fit the data in the region of the $E_{1}$ critical point with a single 2D line shape. For the remaining Mn samples the broadening of $E_{1}$ was large enough that it affected the $E_{0^{'}}$ fit. Therefore for the samples with $x\geq0.028$, the $E_{1}$, $E_{0^{'}}$, \& $E_{2}$ critical points were fit simultaneously. Lastly, as discussed earlier, for samples with $x\geq0.04$ an additional feature could be seen in the derivative spectra (labeled $E_{Mn}$). Therefore in these samples four peaks were fit simultaneously, improving the quality of the fit. As seen in Fig. \ref{fig:fits}, this unfortunately does not provide a good match to this extra peak, therefore the parameters determined for this extra peak are not reported.
	
	The critical point parameters ($E_{g},\Gamma,\Theta$) determined by fitting the numerical second derivative to the form given in eq. \ref{eq:2dcp} are plotted in Fig. \ref{fig:fitparams}. Examining the gap energies plotted in Fig. \ref{fig:fitparams}, we see that the fitting results are in reasonable agreement with our expectations from Figs. \ref{fig:psidel}, \ref{fig:e1e2}, \& \ref{fig:fullderivspec}. Specifically $E_{g}$ of the $E_{1}$ critical point blue shifts with increasing Mn doping, while $E_{0^{'}}$ \& $E_{2}$ remain unchanged within experimental error. In Fig. \ref{fig:fitparams}, we also find that $E_{1}$ critical point is significantly broadened while the other critical points remain mostly unchanged by low-temperature growth. However, it is quite surprising to find that the $E_{1}+\Delta_{1}$ critical point is substantially broadened in LT-GaAs, while only a small increase in the broadening of $E_{1}$ occurs. Finally, in Fig. \ref{fig:fitparams} we see that $\Theta$ for $E_{0^{'}}$ \& $E_2$ appears to grow  as we trace across the samples, but is remains mostly constant for $E_{1}$.
	
\section{Discussion}
\label{sec:discuss}
\subsection{Perturbations of the Critical Point Energies}
	\label{sec:perturb}
	The Hamiltonian of Mn doped GaAs will contain two additional terms due to exchange (Coulomb) and hybridization (Kinetic) between the Mn d orbitals and the As/Ga sp orbitals. The exchange term produces a red shift of the critical points, \cite{exchange,Znelip} yet only blue shifting, if any, is seen in our data. This results from the fact that at room temperature, Ga$_{1-x}$Mn$_x$As is paramagnetic, significantly reducing the effect of the exchange interaction. The effect of sp-d hybridization on the band-gap energies of DMS was first proposed in an ellipsometric study of Cd$_{1-x}$Mn$_x$Te, and has since been described theoretically\cite{hybrid} and observed experimentally in Zn$_{1-x}$(Mn,Fe,Co)$_x$Te\cite{Znelip} and Ga$_{1-x}$Fe$_x$As.\cite{feelip}. Qualitatively the s and p bands of the host are repelled by the d-levels through sp-d hybridization, such that if a d level is above(below) an sp band it pushes the sp band to lower(higher) energy. We note that due to symmetry considerations, hybridization has no effect on the $\Gamma_{6}$, s-like, conduction band at the $\Gamma$ point. However, since this is a second order effect, the shifting is inversely proportional to the energy separating the s,p and d band/level. Carefully examining Fig. \ref{fig:bndstrc}, we expect the separation between the light hole, heavy hole and split-off band to be strongly affected by sp-d hybridization. 
	
	Another term in the Hamiltonian arises from the perturbing potential of the impurities in the sample. This effect was first studied in Si\cite{CardonaSi} and later in Ge\cite{CardonaGe} and GaAs\cite{Cardona3} and agrees well with the result of second order perturbation theory. The impurities, acceptors and/or donors, provide scattering centers such that the self energy is altered. The self energy of a particle in state $|k,n>$ is perturbed by a second order process, whereby it scatters into a virtual intermediate state $|k+q,n'>$ and then back into the original state $|k,n>$.  This results in red shifting and broadening of the critical points. 
	
	If we assume Thomas-Fermi screening, to second order the changes in $E_{g}$ can be written as:
	 \begin{equation}
	\label{eq:perturb}
	\Delta E^{x}_{g}=E^{x}_{g}-E^{0}_{g} \approx \sum_{q} \frac{N_{imp}}{(q^{2}+q_{TF}^{2})}- \sum_{q} \frac{N_{imp}}{(q^{2}+q_{TF}^{2})^2} + x  \sum_{i}  [\frac{V_{(s,p)d}^{2}}{E^{C}-E^{d}_{i}}-\frac{V_{pd}^{2}}{E^{V}-E^{d}_{i}}]+\Delta E^{x}_{Strain}	\end{equation}	
where $E^{x}_{g}$ is the value of the gap at x doping of Mn, $N_{imp}$ is the impurity density, $E^{C,V}$ the energy of the conduction(valence) band involved in the transition,  $E^{d}_{i}$ the energy of the ith Mn level, and $q_{TF}^{2}\propto p^{1/3} m^{*}$ is the Thomas-Fermi wavevector with p the carrier concentration and $m^{*}$ their effective mass. The first and second terms in eq. \ref{eq:perturb} are the first and second order perturbations of the impurity potential.\cite{Cardona3} The first term in eq. \ref{eq:perturb} is generally small and has a different sign for acceptors and donors, such that in heavily compensated materials this term can be neglected. The second order term produces red shifts proportional to $N^{\alpha}_{imp}$, where $\alpha=1(1/3)$ for large(small) q scattering.  For Ga$_{1-x}$Mn$_{x}$As the impurity density is quite large, we therefore expect large q scattering to dominate. The third term, $\Delta E^{x}_{Strain}$ is the shift in the critical point energy due to compressive strain in the thin film. Since the lattice constant of Ga$_{1-x}$Mn$_{x}$As generally follows Vegards law (ie:grows linearly with x), the films will be under increasing compressive strain. As we demonstrate in sec. \ref{sec:E1}, the strain results in a small red shift. The fourth term in eq. \ref{eq:perturb} is the result of second order perturbation theory of the sp-d hybridization.\cite{Hass} Effectively this term results in the sp band "avoiding" the d level, such that if the level is below(above) the sp-band the band will move up(down) in energy. Additionally, although not explicitly stated in eq. \ref{eq:perturb}, $V_{(s,p)d}$ has $\overrightarrow{k}$ dependence that results from the directional dependence of the overlap of sp and d orbitals. Therefore the size $\Delta E_{g}$ will depend on the direction in k space of the transition, the carrier effective mass, and the carrier density, and the density of ionized impurities.
 
	\subsection{$E_0$}
	\label{sec:E0}
	The results presented in this paper provide additional insights into the smearing of the band gap of GaAs grown at low temperatures. In our previous studies of these samples we clearly established that this broadening was, in part, the result of transitions either beginning (in the case of n-type LT-GaAs) or ending (in the case of p-type Ga$_{1-x}$Mn$_{x}$As) in the As$_{Ga}$ impurity states.\cite{jasonpapers} However, with the additional information provided by the $\hat{\epsilon}(E>1.5eV)$ we see that this broadening is also the result of a relaxation of the requirement of momentum conservation. As discussed in the previous section, this relaxation is due to the presence of impurities that provide additional scattering mechanisms. Since transitions are no longer required to be direct, states in the valence band that are not at the zone center can contribute to transitions which end at the zone center. Ultimately this results in a broadening of transitions and a transfer of spectral weight from higher energies to lower ones, as is seen in Fig. \ref{fig:e1e2}. We note that a similar result is found in GaAs damaged by Ion-implantation.\cite{Aspnes1} 

	It is interesting to note that this smearing may also, in part, result from sp-d hybridization. From the positions of the Mn levels in Fig. \ref{fig:bndstrc} we expect the light and heavy hole valence bands to be shifted further than the split-off band. This implies that the splitting between these bands ($\Delta_{0}$) will depend on the doping level and the strength of sp-d hybridization. Eq. \ref{eq:perturb} and Fig. \ref{fig:bndstrc} imply $\Delta_{0}\propto -xV^{2}_{pd}$, such that the valence bands will merge at the $\Gamma$ point for $x\approx0.04$ for $V^{2}_{pd}=0.58 eV$ as determined by photoemission in Ref. \onlinecite{Fujimori}. This merging should lead to a smeared band edge, as is seen in Fig. \ref{fig:e1e2}.
	\subsection{$E_1$ \& $E_{1}+\Delta_{1}$}
	\label{sec:E1}
	The $E_1$ \& $E_{1}+\Delta_{1}$ critical points result from the almost parallel nature of the heavy \& light hole valence bands and the $\Gamma_{6}$ conduction band near the $\Lambda$ point (see Fig. \ref{fig:bndstrc}). The blue shifting of $E_{1}$ is quite surprising as these samples contain a large defect concentration. However, in LT-GaAs $E_{1}$ is unperturbed due to the nature of the defects in this sample, namely As$_{Ga}$. Since As$_{Ga}$ are deep double donors, their electrons are very efficient at screening the impurity potential, preventing As$_{Ga}$ from effecting the band structure. Yet in Ga$_{1-x}$Mn$_{x}$As, as x is increased the Fermi level moves closer to the valence band and the material first becomes fully compensated, then p-type.\cite{vanesch,holedensity,Fujimori,jasonpapers,STS} We therefore expect the screening of the potentials to be significantly reduced at low Mn dopings. Then as the number of carriers increases, the effect of the impurities on the band structure should be diminished. As a result the renormilization of the $E_{1}$ critical point will be substantial at low dopings, then flatten out or possibly be reduced as the number of carriers increases.
	
	The significant blue shifting seen in these critical points suggests the impurity perturbations are overcome by a strong $V_{(s,p)d}$ interaction occurring in the 111 direction. This result is not entirely surprising, given the strong hybridization believed to occur between Mn d and As p orbitals.\cite{Fujimori, mathieu, bandstrc} Additionally, regardless of the site of the substitutional Mn atom in the unit cell, it will always have As neighbors in the 111 and/or $\overline{111}$ directions (see Fig. \ref{fig:bndstrc}). To qualitatively evaluate  eq. \ref{eq:perturb} for $E_1$, we must carefully consider the result of adding a 3d$^5$ local moment to the GaAs environment.  Examining Fig. \ref{fig:bndstrc} we see that the d-levels are far in energy from the bands involved in the $E_1$ critical point, such that they would most likely cause a small red shift of this transition. However the Mn acceptor level is just above the GaAs valence bands. Photoemission on Ga$_{1-x}$Mn$_x$As has demonstrated the d-like character of this level as well as it's strong hybridization with the As 3p states.\cite{Fujimori}

\begin{figure}
\includegraphics*[width=5.0in]{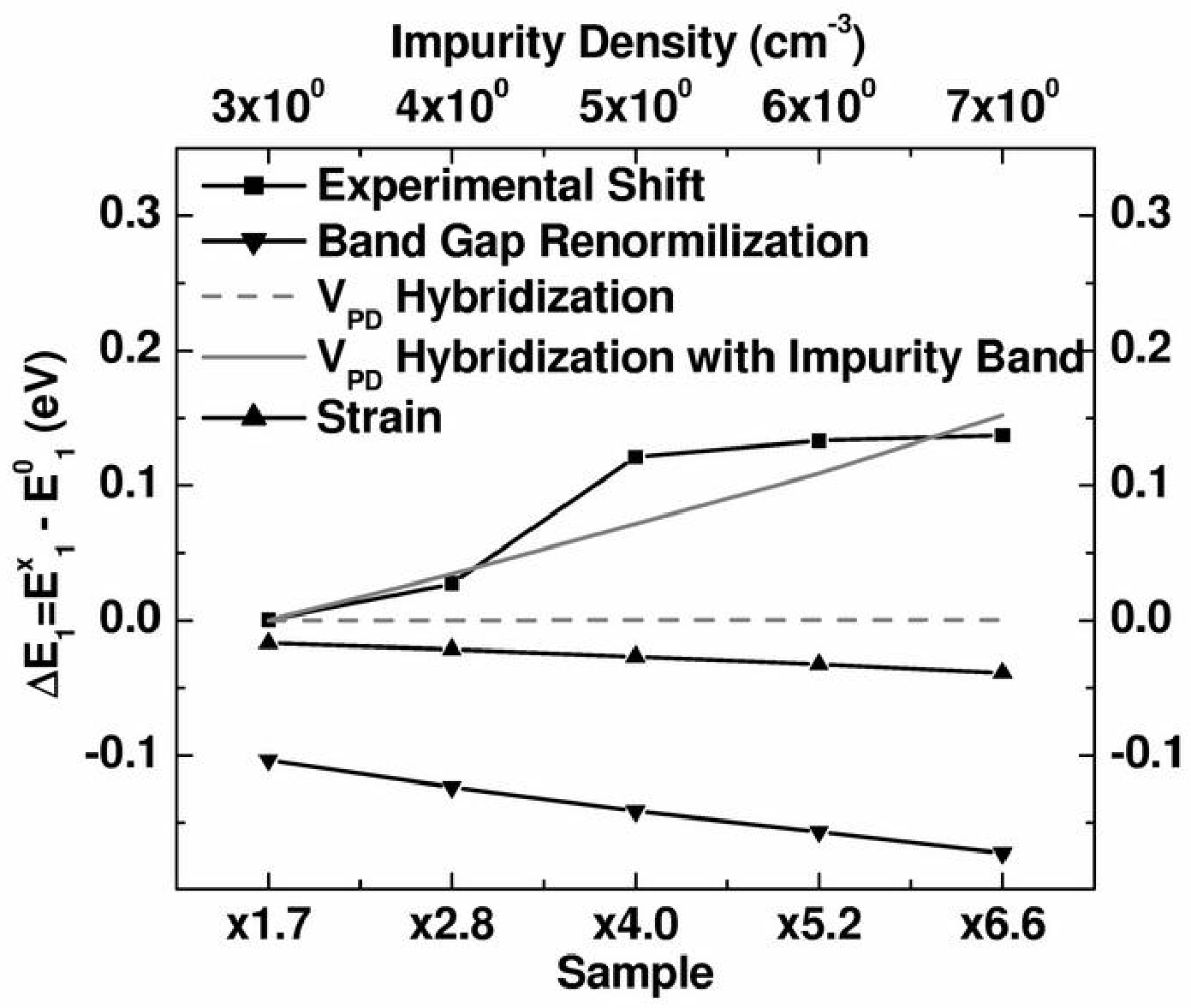}
\caption{\label{fig:E1shift} (Top panel) The measured shift in $E_{1}$ with increasing x. The red shifts due to strain and band gap renormalization are also plotted. The shift of $E_{1}$ resulting from hybridization between the sp and d levels are drawn in gray. The impurity band must clearly be included in the hybridization to explain the blue shift in $E_{1}$. Lines are guides to the eyes.} 
\end{figure}

To quantitatively examine these trends, in Fig. \ref{fig:E1shift} we have plotted $\Delta E^{x}_{1}= E^{x}_{1}- E^{1.7}_{1}$, where $E^{x}_{1}$ is the measured position of the $E_{1}$ critical point at a given doping x. We have chosen to plot the shift this way to account for the merging of the $E_1$ \& $E_{1}+\Delta_{1}$ . Additionally the shifts due to strain, ionized impurities, and pd hybridization. It appears that hybridization between the mn induced impurity band and the GaAs valence band is needed to fully account for the blue shifting. These results also suggest the defects in GaMnAs are well screened by the carriers, which may not be surprising due to their large effective mass.\cite{jasonpapers,Cardona3}. One alternate scenerio, would reduce the separation between the GaAs conduction band and the d$^5$/d$^6$ , such that it lied below the conduction band near the $\Lambda$ point. While this would also result in a blueshifting of E1, we believe this scenerio is highly unlikely, for two reasons. First, from a theoretical standpoint it would require a significant reduction in the $U_{eff}$, which seems highly suspect. Secondly, as discussed in the next section, in higher doped samples we observe evidence of a transition from the valence bands to the d$^5$/d$^6$ level, which agree with it's placement from previous photoemission studies.  We therefore conclude that the blue shifting of $E_1$ with Mn doping is the result of hybridization between the Mn impurity band and the GaAs valence band. As discussed in the pervious section, this hybridization will also reduce the spin-orbit splittings $\Delta_{0} \& \Delta_{1}$, such that is is partially responsible for the apparent merging of the $E_1$ \& $E_{1}+\Delta_{1}$ critical points in Ga$_{1-x}$Mn$_{x}$As. However, since the broadening of these critical points is significantly increased upon mn doping, $E_1$ \& $E_{1}+\Delta_{1}$ cannot be separately distinguished since $\Gamma \cong \Delta_{1}$. Therefore the reduction of  $\Delta_{1}$ cannot be quantitatively accessed using this data set.

	As discussed in Sec. \ref{sec:perturb}, the internal strain in Ga$_{1-x}$Mn$_{x}$As will also result in a red shift of the $E_{1}$ and $E_{1}+\Delta_{1}$ critical points. Using the lattice parameters established in Ref. \onlinecite{ohno} we have estimated the red shift in $E_{1}\leq 0.019$ and $E_{1}+\Delta_{1}\leq0.013$ (see Fig. \ref{fig:E1shift}.\cite{strain}  Additionally these samples are 500nm thick and grown on 60nm buffer layers such that the top most layers of the films should be relaxed. In the vicinity of the $E_{1}$ critical point, $\delta$ is as long as 20 nm. We therefore  conclude strain has little or no effect on measured critical point energies. This also suggests that the broadening of $E_1$ and $E_{1}+\Delta_{1}$ is not the result of a lifting of the degeneracy of the "z" component of angular momentum in the light and heavy hole valence bands. In particular, since $j_{z}=\pm 3/2$ the internal splitting due to strain is more significant for the heavy hole band, therefore the broadening of the $E_{1}$ critical point should be greater than that of the $E_{1}+\Delta_{1}$ critical point. However in LT-GaAs the opposite is observed. Nonetheless the broadening of $E_{1}$ with Mn doping is not surprising given the large number of impurities in these samples, and the resulting relaxation of momentum conservation. Assuming $\Gamma$ follows the trends previously established for doped GaAs,\cite{Cardona3} we expect $\Gamma\cong100meV$ for $E_{1}$ and $E_{1}+\Delta_{1}$, which should grow with increasing impurity concentration. This is qualitatively consistent with our findings of a combined broadening of 220 meV (see Fig. \ref{fig:fitparams}); however a quantitative comparison is not possible due to the uncertainty in carrier and impurity concentrations. 

	\subsection{$E_{0^{'}}$}
	\label{sec:E0'}
	The $E_{0^{'}}$ critical point occurs at the zone center as a result of transitions from the heavy and light hole valence bands to the $\Gamma_{7}~\&~\Gamma_{8}$ conduction bands (see Fig. \ref{fig:bndstrc}). Therefore the $E_{0^{'}}$ critical point provides insight into changes in the electronic structure near the zone center. Given our experimental resolution and fitting methods, we determined the shift in $E_{0^{'}}\leq\pm20$~meV. Given the strong blue shifting seen in the $E_{1}$ critical point, this is quite surprising. Additionally, due to the close proximity of the Mn d$^5$/d$^6$ level to the $\Gamma_{7}~\&~\Gamma_{8}$ conduction bands, see Fig. \ref{fig:bndstrc}, we expect significant blue shifting of $E_{0^{'}}$ from sp-d hybridization. The Mn acceptor level is also quite close to the light and heavy hole valence bands at the $\Gamma$ point. However this apparent null result, can be explained by a reduction in the strength of $V_{(s,p)d}$ at the zone center. We therefore conclude that the hybridization shifts at the zone center are approximately equal to the strength of the re-normalization of the gap from the impurity potentials. It is also interesting to note that the existence of this feature in all mn doped samples, suggests the the Fermi level is less than 200 meV into the GaAs valence band. 
	
		The Mn d$^5$/d$^6$ level also produces another interesting effect on the derivative spectra of Ga$_{1-x}$Mn$_{x}$As. As mentioned in Sec. \ref{sec:cpa}, samples with $x\geq0.04$ contain an extremely weak extra feature, labeled $E_{Mn}$, just below $E_{0^{'}}$. Due to the limited amplitude of this component of $ \frac{d^{2}\epsilon}{dE^{2}}$, it is difficult to discuss in detail. However it's origin may be related to a transition from the valence band to the d$^5$/d$^6$ level (see Fig. \ref{fig:bndstrc}). Similar transitions have been observed in Cd$_{1-x}$Mn$_x$Te and Zn$_{1-x}$Co$_{x}$Te.\cite{cdelip,Znelip} The spectral weight associated with these transitions is generally quite small due to the heavy mass of the d-level. Additionally this level will generally be split due to the crystal field, thereby broadening the transition. 
\subsection{$E_2$}
	\label{sec:E2}
	The $E_{2}$ critical point results from the almost parallel nature of the heavy and light hole valence bands and the $\Gamma_{6}$ conduction band near the X point (see Fig. \ref{fig:bndstrc}). We also expect to see shifts in $E_{2}$ as a result of the perturbing potential of the impurities. Nonetheless this critical point is clearly unchanged by low-temperature growth and/or Mn doping. This apparent null result for the $E_2$ critical point may also be explained by the canceling of the impurity and hybridization terms. However, this spectral region is affected by the presence of an oxide layer. Specifically, the additional layer reduces the measured strength of the $E_2$ critical point, yet it will not affect it's position.\cite{Aspnes2} We therefore attribute the apparent random nature of the strength of this transition seen in Fig. \ref{fig:e1e2} to the presence of the oxide layer, which is not fully accounted for in our model. 
	
	Interestingly, both the $E_{0^{'}}$ and $E_{2}$ critical points see an enhancement of $\Theta$ with increased Mn doping. We believe this results from the additional coulomb potentials of the impurities in these materials.  The potential due to defects in Ga$_{1-x}$Mn$_{x}$As will be quite complicated since it originates from both acceptors and donors. In fact, it appears that the defects tend to cluster,\cite{defects} suggesting they produce dipole or higher order fields. These correlated potentials should be much weaker and more complex than the potential of independent impurities. This may also explain the subdued red shifting effect of these potentials. 
		
\section{Summary and Outlook}
	This work is the first ellipsometric study of Ga$_{1-x}$Mn$_{x}$As. In this paper we have clearly detailed the progression of the GaAs band structure upon doping with Mn. The $E_{1}$ transition blue shifts with increasing Mn doping, while all other critical points remain unchanged.  This blue shifting of $E_{1}$ is the result of sp-d hybridization of the Mn induced impurity band and the GaAs valence band. This finding also signals the existence of the Mn induced impurity band throughout the entire doping range. Additionally these measurements support the conclusion that this band has primarily d-character. The fact that blueshifting is only seen in the $E_{1}$ critical point indicates the strength of $V_{(s,p)d}$ is larger in the 111 direction.  It is interesting to note that the anisotropy of $V_{(s,p)d}$ seen here likely plays a role in the anisotropic magneto-resistance of Ga$_{1-x}$Mn$_{x}$As.\cite{anisomr}  The significant increase in broadening of the critical points also establishes the relaxation of the conservation of momentum in these materials. However $\overrightarrow{k}$ still appears to be a good quantum number in this system, as the $E_1$, $E_{0^{'}}$, and $E_2$ critical points can all be resolved at every doping level in this study. Additionally the band structure of GaAs appears to remain mostly intact, despite the large number of defects found in these materials. However these results also suggest a significant reduction in spin-orbit splitting in Ga$_{1-x}$Mn$_{x}$As.If confirmed, the reduction in spin-orbit splitting implies long spin lifetimes, making Ga$_{1-x}$Mn$_{x}$As an excellent candidate for spintronic materials. Interestingly, the band gap renormilization due to defects  is compensated by sp-d hybridization. Furthermore these results imply sp-d hybridization plays a key role in the optical properties of Ga$_{1-x}$Mn$_{x}$As.
		
	Key insights into the Hamiltonian governing Ga$_{1-x}$Mn$_{x}$As are clearly provided by this work. Specifically, it is clear that the Mn impurity band plays an important role at all doping levels. Additionally $\overrightarrow{k}$ is partially relaxed in these materials, confirming the assertion that Ga$_{1-x}$Mn$_{x}$As have the electronic structure of an alloy. It is also clear that the impurity potentials are strongly screened in these materials, either by heavy carriers and/or by other impurities. As this is the first ellipsometric study of a fully compensated semiconductor, it is unclear what role defect correlations play in reducing the perturbation of impurity potentials on the band structure. Therefore further theoretical and experimental evaluation of this problem is clearly called for. However the defects and additional impurity states in these materials result in a large broadening of the critical points. Therefore low temperature measurements are needed to help resolve the exact position of the critical points and the magnitude of spin-orbit splitting. Additionally the effect of electron-phonon coupling and potentially the position of d$^5$/d$^6$ level could be determined with temperature dependent ellipsometry. Nonetheless this study provides a unique litnus test for further calculations of the Ga$_{1-x}$Mn$_x$As band structure. In fact, one of the reasons the GaAs band structure is so well understood is the large number of calculations based upon and/or compared to experimental determinations of its critical points. We therefore believe these results will be critical in determining the physics governing Ga$_{1-x}$Mn$_{x}$As.

\begin{acknowledgments}
This work was supported by the DOE, NSF, DARPA, and ONR. We are grateful for numerous discussions with L. Cywinski, M. Fogler, A. Fujimori, E.M. Hankiewicz, J. McGuire, L.J. Sham, J. Sinova and T. Tiwald.
\end{acknowledgments}

\clearpage

\begin{table}

\caption{\label{TBL}Parameters of the samples studied, which were grown at a substrate temperature of 265~C, with As/Mn beam flux ratio of $\sim$200/1. Ga growth rates were $\sim$0.3 ML/s and Mn growth rates were 0.02-0.05 ML/s. All thicknesses are in nm and T$_{C}$ are in Kelvin.}
\begin{ruledtabular}

\begin{tabular}{cccccc}
Sample&
Surface Layer&
Oxide Layer&
Generic Layer&
$T_{C}$
\tabularnewline
\hline
\hline 
GaAs&
0.211&
2.966&
n.a.&
$n.a.$
\tabularnewline
\hline 
LT-GaAs&
0.289&
4.64&
1558.5&
$n.a.$
\tabularnewline
\hline 
Ga$_{0.983}$Mn$_{0.017}$As&
0.332&
3.973&
514.47&
$<5$
\tabularnewline
\hline 
Ga$_{0.072}$Mn$_{0.028}$As&
0.846&
3.317&
480.17&
$30$
\tabularnewline
\hline 
Ga$_{0.060}$Mn$_{0.040}$As&
0.848&
2.533&
485.47&
$45$
\tabularnewline
\hline 
Ga$_{0.048}$Mn$_{0.052}$As&
0.918&
4.075&
479.57&
$70$
\tabularnewline
\hline 
Ga$_{0.034}$Mn$_{0.066}$As&
0.88&
3.138&
497.96&
$70$
\tabularnewline

\end{tabular}

\end{ruledtabular}

\end{table}

\clearpage

\begin{thebibliography}
\bibitem{backsub} we note that intereference resulting from reflection off of the back of the substrate is possible. However these generally are incoherent and therefore reduce the reliability of the model. Therefore we have developed a technique to roughen the back surface of the substrate without damaging the films to prevent this interference from occuring.
\end{thebibliography}
\end{document}